\documentstyle[aps,twocolumn,epsfig]{revtex}

\begin{document}
\draft

\twocolumn[\hsize\textwidth\columnwidth\hsize\csname@twocolumnfalse\endcsname
\title{Quantum teleportation of entangled coherent states}
\author{Xiaoguang Wang}
\address{Institute of Physics and Astronomy, Aarhus University, \\
DK-8000, Aarhus C, Denmark}
\date{\today}
\maketitle

\begin{abstract}
We propose a simple scheme for the quantum teleportation of both bipartite 
and multipartite entangled
coherent states with the successful probability 1/2. The scheme is based on  
only linear optical devices such as  beam splitters and phase shifters, 
and two-mode photon number measurements. The quantum channels described 
by the multipartite maximally entangled coherent states are readily made by the beam
splitters and phase shifters.
\end{abstract}

\pacs{PACS numbers: 03.67.Hk, 03.65.Ud, 03.67.Lx}

] \narrowtext

Quantum teleportation, first proposed by Bennett {\it et al}. \cite{Tele},
is a disembodied transport of quantum states between subsystems through a
classical communication channel requiring a shared entangled state. Several
experiments have been implemented to demonstrate the teleportation\cite{Exp}%
. Most of the studies have been confined to the teleportation of single-body
quantum states: quantum teleportation of two-level states\cite{Tele}, $N$%
-dimensional states\cite{Telequnit}, and continuous variables\cite{Telecv}.
Recently Lee and Kim considered the teleportation of bipartite entangled
states through noisy quantum channels\cite{Leekim}. Ikram {\it et al}. \cite
{Ikram} and Shi {\it et al}.\cite{Shi} proposed schemes for the quantum
teleportation of a two-qubit entangled state. Teleportation of some pure
entangled states of both discrete and continuous variables is considered by
Gorbachev {\it et al}.\cite{Gorbachev}. A possibility to copy pure entangled
states was studied by Koashi and Imoto\cite{Koashi}.

In the teleportation schemes we need certain types of maximally entangled
states (MES). We consider the following entangled coherent states (ECS)\cite
{Barry}

\begin{eqnarray}
|\alpha ;\alpha \rangle _{12}^{\pm } &=&\frac 1{\sqrt{2(1\pm e^{-4|\alpha
|^2})}}(|\alpha \rangle _1|\alpha \rangle _2  \nonumber \\
&&\pm |-\alpha \rangle _1|-\alpha \rangle _2),  \label{eq:aa}
\end{eqnarray}
where $|\alpha \rangle _i$ ($i=1,2$) is the coherent state of system $i$. It
is interesting to see that the ECS $|\alpha ;\alpha \rangle _{12}^{-}$ is a
MES, irrespective of the parameter $\alpha $\cite{vanEnk}. The ECS $|\alpha
;\alpha \rangle _{12}^{-}\,$can be rewritten as the form

\begin{equation}
|\alpha ;\alpha \rangle _{12}^{-}=\frac 1{\sqrt{2}}\left( |\alpha \rangle
_1^{+}|\alpha \rangle _2^{-}+|\alpha \rangle _1^{-}|\alpha \rangle
_2^{+}\right)   \label{eq:alpha}
\end{equation}
in terms of the even and odd coherent states 
\begin{equation}
|\alpha \rangle _i^{\pm }=\frac 1{\sqrt{2(1\pm e^{-2|\alpha |^2})}}\left(
|\alpha \rangle _i\pm |-\alpha \rangle _i\right) .
\end{equation}
Eq.(\ref{eq:alpha}) shows that the state $|\alpha ;\alpha \rangle
_{12}^{-}\,\,$manifestly has one ebit of entanglement. In the limit $|\alpha
|\rightarrow 0,$ the ECS reduces to the singlet-like state $|\Psi
^{+}\rangle _{12}=\left( |0\rangle _1|1\rangle _2+|1\rangle _1|0\rangle
_2\right) /\sqrt{2},$ where $|0\rangle _i$ and $|1\rangle _i$ are photon
number states (Fock states).

van Enk and Hirota\cite{vanEnk} have discussed how to teleport a
Schr\"{o}dinger cat state of the form

\begin{eqnarray}
|\alpha \rangle _{\text{cat}} &=&{\cal N}(\epsilon _{+}|\alpha \rangle
+\epsilon _{-}|-\alpha \rangle ),  \nonumber \\
{\cal N} &=&\left[ |\epsilon _{+}|^2+|\epsilon _{-}|^2+2e^{-2|\alpha |^2}%
\mathop{\rm Re}%
(\epsilon _{+}\epsilon _{-}^{*})\right] ^{-1/2},
\end{eqnarray}
through a quantum channel described by the MES $|\alpha ;\alpha \rangle
_{12}^{-},$ where $\epsilon _{\pm }$ are complex numbers. Inspired by their
teleportation scheme, we consider the teleportation of the following ECS

\begin{eqnarray}
|\Phi \rangle _{12} &=&{\cal N}_\Phi (\epsilon _{+}|\alpha \rangle _1|\alpha
\rangle _2+\epsilon _{-}|-\alpha \rangle _1|-\alpha \rangle _2),  \nonumber
\\
{\cal N}_\Phi &=&\left[ |\epsilon _{+}|^2+|\epsilon _{-}|^2+2e^{-4|\alpha
|^2}%
\mathop{\rm Re}%
(\epsilon _{+}\epsilon _{-}^{*})\right] ^{-1/2}.
\end{eqnarray}

In the teleportation of entangled states, particularly two-qubit pure
states, one can use two EPR pairs, a four-qubit quantum channel, or a less
expensive three-qubit GHZ state\cite{Gorbachev}. If we want to teleport the
ECS $|\Phi \rangle _{12},$ we need at least a tripartite entangled state as
the quantum channel. In a recent paper\cite{Wang}, we have considered the
following tripartite entangled states

\begin{eqnarray}
|\sqrt{2}\alpha ;\alpha ;\alpha \rangle _{345}^{\pm } &=&\frac 1{\sqrt{%
2(1\pm e^{-8|\alpha |^2})}}(|\sqrt{2}\alpha \rangle _3|\alpha \rangle
_4|\alpha \rangle _5  \nonumber \\
&&\pm |-\sqrt{2}\alpha \rangle _3|-\alpha \rangle _4|-\alpha \rangle _5).
\label{eq:pm}
\end{eqnarray}
The bipartite entanglement of the tripartite states can be characterized by
one measure of entanglement, the concurrence\cite{Concurrence}. The
concurrence of the state $|\sqrt{2}\alpha ;\alpha ;\alpha \rangle
_{345}^{\pm }\,$between system $i$ and systems $j,k$ $(i\neq j\neq k\in
\{3,4,5\})$ is denoted by $C_{i(jk)}^{\pm }.$ The concurrences are obtained
as\cite{Wang} 
\begin{eqnarray}
C_{3(45)}^{+} &=&\tanh (4|\alpha |^2),\text{ }C_{3(45)}^{-}=1,  \nonumber \\
C_{4(35)}^{\pm } &=&C_{5(34)}^{\pm }=\frac{\sqrt{(1-e^{-4|\alpha
|^2})(1-e^{-12|\alpha |^2})}}{1\pm e^{-8|\alpha |^2}}.  \label{eq:cc}
\end{eqnarray}
$\,$We see that the system 3 with systems 4,5 is always maximally entangled
in the state $|\sqrt{2}\alpha ;\alpha ;\alpha \rangle _{345}^{-}$. This
tripartite state may be considered as a tripartite extension of the
bipartite MES $|\alpha ;\alpha \rangle _{12}^{-}$ and will act as the
quantum channel in the following discussions. Now having the state $|\Phi
\rangle _{12}$ to be teleported and the MES $|\sqrt{2}\alpha ;\alpha ;\alpha
\rangle _{345}^{-}$ as a quantum channel, we begin to discuss our
teleportation scheme.

We first briefly review the action of a beam splitter on coherent states.
The lossless symmetric 50/50 beam splitter is described by $B_{12}=e^{i\frac %
\pi 4(a_1^{\dagger }a_2+a_2^{\dagger }a_1)},$ which transforms the coherent
states $|\alpha \rangle _1|\beta \rangle _2$ as

\begin{equation}
B_{12}|\alpha \rangle _1|\beta \rangle _2=|(\alpha +i\beta )/\sqrt{2}\rangle
_1|(\beta +i\alpha )/\sqrt{2}\rangle _2.
\end{equation}
Here $a_i$ and $a_i^{\dagger }$ are the bosonic annihilation and creation
operators of system $i$, respectively. By equipping the beam splitter by a
pair of $-\pi /2$ phase shifters described by the unitary operator $%
P_2=e^{-i\pi a_2^{\dagger }a_2/2}$ , we can have the operator ${\cal B}%
_{12}=P_2B_{12}P_2$ which transforms the state $|\alpha \rangle _1|\beta
\rangle _2$ as

\begin{equation}
{\cal B}_{12}|\alpha \rangle _1|\beta \rangle _2=|(\alpha +\beta )/\sqrt{2}%
\rangle _1|(\alpha -\beta )/\sqrt{2}\rangle _2.  \label{eq:bb}
\end{equation}
This transformation plays a key role in our teleportation scheme.

Now Alice wish to teleport the ECS $|\Phi \rangle _{12}$ to the remote
partner Bob by sharing the MES $|\sqrt{2}\alpha ;\alpha ;\alpha \rangle
_{345}^{-}.$ The systems 1,2 and 3 are at Alice's side and systems 4 and 5
are at Bob' side. The initial state of the whole system is then given by

\begin{equation}
|\Psi \rangle _{12345}=|\Phi \rangle _{12}|\sqrt{2}\alpha ;\alpha ;\alpha
\rangle _{345}^{-}
\end{equation}

We first apply the transformation ${\cal B}_{21}=P_1B_{21}P_1$ to the
initial state. From Eq.(\ref{eq:bb}), the state after the transformation
becomes a direct product of the vacuum state $|0\rangle _1$ with the
unnormalized state

\begin{eqnarray}
|\Psi ^{\prime }\rangle _{2345} &=&\epsilon _{+}(|\sqrt{2}\alpha \rangle _2|%
\sqrt{2}\alpha \rangle _3|\alpha \rangle _4\otimes |\alpha \rangle _5 
\nonumber \\
&&-|\sqrt{2}\alpha \rangle _2|-\sqrt{2}\alpha \rangle _3|-\alpha \rangle
_4\otimes |-\alpha \rangle _5)  \nonumber \\
&&+\epsilon _{-}(|-\sqrt{2}\alpha \rangle _2|\sqrt{2}\alpha \rangle
_3|\alpha \rangle _4\otimes |\alpha \rangle _5  \nonumber \\
&&-|-\sqrt{2}\alpha \rangle _2|-\sqrt{2}\alpha \rangle _3|-\alpha \rangle
_4\otimes |-\alpha \rangle _5).
\end{eqnarray}
Now the system 1 is separated from the remain systems. Then by applying the
second transformation ${\cal B}_{23},$ we obtain

\begin{eqnarray}
|\Psi ^{\prime \prime }\rangle _{2345} &=&{\cal B}_{23}|\Psi ^{\prime
}\rangle _{2345}  \nonumber \\
&=&\epsilon _{+}(|2\alpha \rangle _2|0\rangle _3|\alpha \rangle _4\otimes
|\alpha \rangle _5  \nonumber \\
&&-|0\rangle _2|2\alpha \rangle _3|-\alpha \rangle _4\otimes |-\alpha
\rangle _5)  \nonumber \\
&&-\epsilon _{-}(|-2\alpha \rangle _2|0\rangle _3|-\alpha \rangle _4\otimes
|-\alpha \rangle _5  \nonumber \\
&&-|0\rangle _2|-2\alpha \rangle _3|\alpha \rangle _4\otimes |\alpha \rangle
_5).  \label{eq:psi}
\end{eqnarray}
After these two transformations Alice performs a two-mode number measurement
on the modes 2 and 3. The probability of finding $n$ and $m$ photons in
modes 2 and 3 is given by

\begin{equation}
P(n,m)=|_2\langle n|_3\langle m|\Psi ^{\prime \prime }\rangle _{2345}|^2.
\label{eq:p}
\end{equation}
The probability is zero if both $n$ and $m$ are nonzero, i.e., one of the
two integers must be zero in order to have nonzero probability. 

Let us suppose $n\neq 0$ and $m=0$. In this case the state on Bob' side
collapses into

\begin{equation}
|\Phi ^{\prime }\rangle _{45}=\epsilon _{+}|\alpha \rangle _4|\alpha \rangle
_5-\epsilon _{-}(-1)^n|-\alpha \rangle _4|-\alpha \rangle _5,
\end{equation}
For the case $n=0$ and $m\neq 0$ , the state on Bob' side collapses into

\begin{equation}
|\Phi ^{\prime \prime }\rangle _{45}=\epsilon _{+}|-\alpha \rangle
_4|-\alpha \rangle _5-\epsilon _{-}(-1)^m|\alpha \rangle _4|\alpha \rangle
_5.
\end{equation}
Now Alice sends a classical information to Bob and Bob makes a local
transformation $(-1)^{a_4^{\dagger }a_4+a_5^{\dagger }a_5}$ on his state $%
|\Phi ^{\prime \prime }\rangle _{45}.\,$The local transformation is a
multiplication of two $\pi $ phase shifters of modes 4 and 5 and the
resultant state after the transformation is just the state $|\Phi ^{\prime
}\rangle _{45}.$

We see that provided $n$ is odd, the teleportation scheme works perfectly.
However for even $n,$ the transformation for perfect teleportation is

\begin{mathletters}
\begin{eqnarray}
|\alpha \rangle _4|\alpha \rangle _5 &\rightarrow &|\alpha \rangle _4|\alpha
\rangle _5, \\
|-\alpha \rangle _4|-\alpha \rangle _5 &\rightarrow &-|-\alpha \rangle
_4|-\alpha \rangle _5,
\end{eqnarray}
which is in general not a unitary transformation except the limit case $%
|\alpha |\rightarrow \infty .$ From Eqs.(\ref{eq:psi}) and (\ref{eq:p}), the
probabilities $P(0,n)$ and $P(n,0)$ for odd $n$ are obtained as

\end{mathletters}
\begin{equation}
P(0,n)=P(n,0)=\frac{e^{-4|\alpha |^2}|2\alpha |^{2n}}{2n!(1-e^{-8|\alpha
|^2})}.
\end{equation}
Then the probability of success is given by

\begin{equation}
P_{odd}=2\sum_{\text{odd }n}P(n,0)=\frac 12.  \label{eq:odd}
\end{equation}

As seen from Eq.(\ref{eq:odd}), the successful probability is independent on
both $\alpha $ and $\epsilon _{\pm }.$ One restriction to the ECS to be
teleportated is that the mean photon number of the coherent state $|\alpha
\rangle $($|\alpha |^2$) in the system 1(2) is the same as that in the
system 4(5). In other words the state to be teleportated is strongly
correlated in photon number to the quantum channel. Actually the scheme is
used to teleport a qubit encoded in the ECS $|\Phi \rangle _{12}.$ The
teleportation scheme is not optimal, however indeed it gives the nonzero
probability 1/2 independent of $\alpha $.

There is another problem left that if we can produce the tripartite
maximally entangled coherent state which plays the role of the quantum
channel. If we can not, the scheme does not work. Fortunately we can create
this MES in an easy way. We first prepare the systems 3, 4, and 5 in the
state $|2\alpha \rangle _3^{-}|0\rangle _4|0\rangle _5.$ Then by applying
the transformation ${\cal B}_{45}{\cal B}_{34},$ we obtain

\begin{eqnarray}
&&{\cal B}_{45}{\cal B}_{34}|2\alpha \rangle _3^{-}|0\rangle _4|0\rangle _5 
\nonumber \\
&=&{\cal B}_{45}(|\sqrt{2}\alpha \rangle _3|\sqrt{2}\alpha \rangle _4\otimes
|0\rangle _5  \nonumber \\
&&-|-\sqrt{2}\alpha \rangle _3|-\sqrt{2}\alpha \rangle _4\otimes |0\rangle
_5)/\sqrt{2(1-e^{-8|\alpha |^2})}  \nonumber \\
&=&|\sqrt{2}\alpha ;\alpha ;\alpha \rangle _{345}^{-},
\end{eqnarray}
which is just the tripartite MES in the teleportation scheme. In a short
summary we can let the initial state of the whole composite system be $|\Phi
\rangle _{12}|2\alpha \rangle _3^{-}|0\rangle _4|0\rangle _5.$ Then apply a
transformation ${\cal B}_{23}{\cal B}_{21}{\cal B}_{45}{\cal B}_{34}$ to the
initial state, and make the two-mode number measurement to implement the
teleportation.

We would like to investigate further what the successful probability is if
we use a nonmaximally entangled state as the quantum channel in the
teleportation scheme. We choose the state as $|\sqrt{2}\alpha ;\alpha
;\alpha \rangle _{345}^{+}$(\ref{eq:pm}). From Eq.(\ref{eq:cc}) we see that
the state $|\sqrt{2}\alpha ;\alpha ;\alpha \rangle _{345}^{+}$ is not
maximally entangled except the limit case $|\alpha |\rightarrow \infty .$
The entangled state $|\sqrt{2}\alpha ;\alpha ;\alpha \rangle _{345}^{+}$ can
be generated similarly as the state $|\sqrt{2}\alpha ;\alpha ;\alpha \rangle
_{345}^{-}.$

Following the same steps as before, after Alice measures $n$ photons in mode
2 and zero photons in mode 3, Bob's state collapses into the state

\begin{equation}
|\tilde{\Phi}^{\prime }\rangle _{45}=\epsilon _{+}|\alpha \rangle _4|\alpha
\rangle _5+\epsilon _{-}(-1)^n|-\alpha \rangle _4|-\alpha \rangle _5,
\end{equation}
The perfect teleportation is obtained for even $n.$ The successful
probability is given by

\begin{eqnarray}
P_{even} &=&\sum_{\text{even }n>0}[P(0,n)+P(n,0)]  \nonumber \\
&=&\frac{(1-e^{-4|\alpha |^2})^2}{2(1+e^{-8|\alpha |^2})},
\end{eqnarray}
which is independent on the parameters $\epsilon _{\pm }.$ However it
depends on the parameter 
\mbox{$\vert$}%
$\alpha |.$ In the limit 
\mbox{$\vert$}%
$\alpha |\rightarrow \infty ,$ the successful probability becomes 1/2. In
this limit case, the amount of entanglement in the state $|\sqrt{2}\alpha
;\alpha ;\alpha \rangle _{345}^{+}$ is one ebit and the probability can be
1/2. Except this case, the successful probability is always less than 1/2 as
the corresponding quantum channel is not a MES.

Our teleportation scheme can be generalized to the multipartite cases. For
the sake of simplicity we only study the tripartite case. We consider a
tripartite ECS,

\begin{equation}
|\Phi \rangle _{123}=\epsilon _{+}|\sqrt{2}\alpha \rangle _1|\alpha \rangle
_2|\alpha \rangle _3+\epsilon _{-}|-\sqrt{2}\alpha \rangle _1|-\alpha
\rangle _2|-\alpha \rangle _3.
\end{equation}
To teleport $|\Phi \rangle _{123},$ we may need the four-particle entangled
state

\begin{eqnarray}
&&|2\alpha ;\sqrt{2}\alpha ;\alpha ;2\alpha \rangle _{4567}^{-}  \nonumber \\
&=&|2\alpha \rangle _4|\sqrt{2}\alpha \rangle _5|\alpha \rangle _6|\alpha
\rangle _7  \nonumber \\
&&-|-2\alpha \rangle _4|-\sqrt{2}\alpha \rangle _5|-\alpha \rangle
_6|-\alpha \rangle _7,
\end{eqnarray}
which is maximally entangled between the system 4 and systems 5, 6, and 7%
\cite{Wang}. First we disentangle the system 3 from systems 1 and 2 by
applying the operator ${\cal B}_{31}{\cal B}_{32}$ to the state $|\Phi
\rangle _{123}.$ We obtain

\begin{equation}
{\cal B}_{31}{\cal B}_{32}|\Phi \rangle _{123}=|0\rangle _1|0\rangle
_2(\epsilon _{+}|2\alpha \rangle _3+\epsilon _{-}|-2\alpha \rangle _3).
\end{equation}
Then we make the transformation ${\cal B}_{34},$ the two-mode number
measurement on modes 3 and 4, and a classical communication from Alice to
Bob to finish the teleportation process. The successful probability is also
1/2. The quantum channel described by the four-particle entangled state can
be obtained as

\begin{eqnarray}
&&|2\alpha ;\sqrt{2}\alpha ;\alpha ;2\alpha \rangle _{4567}^{-}  \nonumber \\
&=&{\cal B}_{67}{\cal B}_{56}{\cal B}_{45}(|2\sqrt{2}\alpha \rangle _4-|-2%
\sqrt{2}\alpha \rangle _4)|0\rangle _5|0\rangle _6|0\rangle _7.
\end{eqnarray}
It is straightforward to generalize the teleportation scheme to teleport the
multipartite (more than three) entangled ECS of the form $|\Phi \rangle
_{123}$.

In the teleportation scheme described above the probability of success is
1/2 and independent of $\alpha .$ The it is interesting to consider the
limit $|\alpha |\rightarrow 0.$ The state $|\sqrt{2}\alpha ;\alpha ;\alpha
\rangle _{345}^{-}$ can be rewritten as

\begin{eqnarray}
&&|\sqrt{2}\alpha ;\alpha ;\alpha \rangle _{345}^{-}  \nonumber \\
&=&\frac 1{\sqrt{2}}(|\sqrt{2}\alpha \rangle _3^{-}|\alpha ;\alpha \rangle
_{45}^{+}+|\sqrt{2}\alpha \rangle _3^{+}|\alpha ;\alpha \rangle _{45}^{-}),
\end{eqnarray}
which directly leads to

\begin{eqnarray}
|\psi \rangle _{345} &=&\lim_{|\alpha |\rightarrow 0}|\sqrt{2}\alpha ;\alpha
;\alpha \rangle _{345}^{-}  \nonumber \\
&=&\frac 1{\sqrt{2}}(|1\rangle _3|00\rangle _{45}+|0\rangle _3|\Psi
^{+}\rangle _{45}).
\end{eqnarray}
Here the state $|00\rangle _{45}\equiv |0\rangle _4|0\rangle _5.$ We use the
state $|\psi \rangle _{345}$ to teleport the state 
\begin{equation}
|\phi \rangle _{12}=\frac 1{\sqrt{|a|^2+|b|^2}}(a|00\rangle _{12}+b|\Psi
^{+}\rangle _{12}).
\end{equation}
Then the initial state of the whole system is given by 
\begin{equation}
|\psi ^{\prime }\rangle _{12345}=|\phi \rangle _{12}|\psi \rangle _{345}.
\end{equation}
$\,$It is straightforward to check that 
\begin{mathletters}
\begin{eqnarray}
{\cal B}_{12}|10\rangle _{12} &=&|\Psi ^{+}\rangle _{12},\,{\cal B}%
_{12}|01\rangle _{12}=|\Psi ^{-}\rangle _{12},  \label{eq:bb1} \\
{\cal B}_{12}|\Psi ^{+}\rangle _{12} &=&|10\rangle _{12},\text{ }{\cal B}%
_{12}|\Psi ^{-}\rangle _{12}=|01\rangle _{12}.  \label{eq:bb2}
\end{eqnarray}
where $|\Psi ^{\pm }\rangle _{12}=(|10\rangle \pm |01\rangle )/\sqrt{2}.$ We
have used the identity ${\cal B}_{12}^2=1.$ Then by applying the operator $%
{\cal B}_{13}{\cal B}_{12}$ to the initial state, we obtain

\end{mathletters}
\begin{eqnarray}
{\cal B}_{13}{\cal B}_{12}|\psi ^{\prime }\rangle _{12345} &=&\frac 1{2\sqrt{%
|a|^2+|b|^2}}|0\rangle _2  \nonumber \\
&&[|10\rangle _{13}(a|00\rangle _{45}+b|\Psi ^{+}\rangle _{45})  \nonumber \\
&&+|01\rangle _{13}(-a|00\rangle _{45}+b|\Psi ^{+}\rangle _{45})  \nonumber
\\
&&+a|00\rangle _{13}|\Psi ^{+}\rangle _{45}  \nonumber \\
&&+\frac b{\sqrt{2}}(|20\rangle _{13}-|02\rangle _{13})|00\rangle _{45}].
\end{eqnarray}
Here we have used Eqs.(\ref{eq:bb1}) and (\ref{eq:bb2}). We see the system 2
decouples from the other systems. The teleportation scheme works perfectly
if the resultant state of the two-mode number measurement is $|10\rangle
_{13}.$ If the resultant state is $|01\rangle _{13},$ then Alice needs to
communicate classically with Bob and Bob makes a local transformation $%
(-1)^{a_4^{\dagger }a_4+a_5^{\dagger }a_5}$ to finish the teleportation.
Again the probability of success is 1/2. The similar setup as this
teleportation scheme has been proposed to perform optical state truncation%
\cite{Pegg}, optical simulation of quantum logic\cite{Cerf}, and the quantum
teleportation\cite{Lee}.

The measurement we have used in our teleportion scheme is the two-mode
number measurement which must be sensitive enough to measure the number of
photons and determine whether the number is even or odd. In practise this is
a difficult part especially for large number of photons, but in principle
this can be done. Here we propose one method to distinguish even and odd $n$
in the Fock state $|n\rangle $ by coupling the field to a two-level atom
through dispersive interaction\cite{D}. The Hamiltonian is given by $%
H=ga^{\dagger }a\sigma _x,$ where $g$ is the coupling constant and $\sigma
_\beta (\beta =x,y,z)$ are the Pauli operators describing the peudospin of
the two-level atom.\thinspace This Hamiltonian can be realized in both
cavity-QED and trapped-ion systems\cite{D}. At time $t=\pi /(2g),\,$the
evolution operator $\exp (-iHt)$ becomes $U=\exp (-i\pi a^{\dagger }a\sigma
_x/2).$ Initially let the atom be in the ground state and the field be in
the Fock state $|n\rangle .$ After applying the unitary operator $U$ we can
know that the photon number is even (odd) if the atom is found to be in the
ground (excited) state$.$ By this technique, we can distinguish even and odd
photon numbers.

In conclusion we have proposed a simple scheme to teleport both the
bipartite and multipartite ECS with the successful probability 1/2,
independent of $\alpha $. As a crucial ingredient in the scheme, the quantum
channel described by the multipartite ECS, can be readily made by only
linear optical devices such as beam splitters and phase shifters. Thus we
provide a way to achieve all linear optical teleportation of quantum states%
\cite{Cerf,Lee}. The probability of success in our scheme is 1/2 due to the
use of only linear operations and the absence of photon-photon interaction%
\cite{Vaidman}. Both the measurement and the preparation of the quantum
channel can be implemented in the experiments by the present techniques. We
expect that the present scheme can be used in the experiments to demonstrate
the quantum teleportation of the entangled states.

\acknowledgments
The author thanks the helpful discussions with Klaus M\o lmer, Anders S\o
rensen, Barry C Sanders, Paolo Zanardi, Serge Massar, and Nicolas J Cerf.
This work is supported by the Information Society Technologies Programme
IST-1999-11053, EQUIP, action line 6-2-1.

\end{document}